\documentclass[11pt]{article}
\newcommand{\bc}{\begin{center}}
\newcommand{\ec}{\end{center}}
\newcommand{\be}{\begin{equation}}
\newcommand{\ee}{\end{equation}}
\newcommand{\bea}{\begin{eqnarray}}
\newcommand{\eea}{\end{eqnarray}}
\newcommand{\ba}{\begin{array}}
\newcommand{\ea}{\end{array}}
\newcommand{\edc}{\end{document}}

\def\O{\Omega}

\def\b{\beta}

\def\s{\sigma}
\def\m{\mu}

\def\Z{\mathbf Z}

\def\1{\mathbf 1}

\def\L{\Lambda}

\def\d{\partial}

\begin{document}
\thispagestyle{empty}

{\bf AN EXAMPLE OF ONE - DIMENSIONAL PHASE TRANSITION }
\vspace{0.3cm}
\begin{center}
{\bf U.A. Rozikov}\\
{\it Institute of Mathematics, 29, F.Hodjaev str., 700143, Tashkent,
 Uzbekistan. \\ e-mail: rozikovu@yandex.ru}
\end{center}
\vspace{0.4cm}

--------------------------------------------------------

{\bf Abstract.} In the paper a one-dimensional model with
nearest - neighbor interactions $I_n, n\in \Z$ and  spin values
$\pm 1$ is considered.
We describe a condition on parametres $I_n$ under which
the phase transition occurs.
In particular, we show that the phase transition occurs if
$I_n\geq |n|, n\in \Z.$ \\[1mm]


---------------------------------------------------------

\section{Introduction}
\large

Interest in phase transition in one-dimensional systems has
gained due to Little's work [8]. Existence or nonexistence depends
markedly upon the model employed. In [11] Van Hove shows (see also
[9, section 5.6.]) a one-dimensional system can not exhibit
a phase transition if the (translation-invariant) forces are of
finite range.

An interesting one-dimensional model considered by
 Baur and Nosanow [1], giving rise to a phase transition with only nearest-
neighbor interactions which has some of the interaction constants
equal to minus infinity. Note that in a particular case
the model which we shall consider
here  is a model with only nearest-neighbor interactions
without assumption of existence a interaction equal to minus infinity
(cf. with [1]).

An other examples of one-dimensional phase transitions for
models with long range interactions   were considered in [3-7], [10].

In the paper we consider the Hamiltonian
$$H(\s)= \sum_{l=(x-1, x): x\in \Z}
I_x\1_{\s(x-1)\ne\s(x)}, \eqno(1)$$
where $\Z=\{...,-1,0,1,...\},$ $\s=\{\s(x)\in \{-1,1\}: x\in \Z\}
\in \O=\{-1,1\}^{\Z},$ and $ I_x\in R$ for any $x\in \Z.$

The goal of this paper is to describe a condition on parameters of
the model (1) under which the phase transition occurs.

\section{ Phase transition}
Let us consider the sequence $\L_n=[-n,n], n=0,1,...$ and denote
$\L_n^c=\Z\setminus \L_n$. Consider boundary condition
$\s^{(+)}_n=\s_{\L_n^c}=\{\s(x)=+ 1: x\in \L_n^c\}.$ The energy $H_n^+(\s)$
of the configuration $\s$ in the presence of boundary condition $\s^{(+)}_n$
is expressed by the formula
$$H^+_n(\s)= \sum_{l=(x-1, x): x\in \L_n}
I_x\1_{\s(x-1)\ne\s(x)}+
I_{-n}\1_{\s(-n)\ne 1}+I_{n+1}\1_{\s(n)\ne 1}. \eqno(2)$$

The Gibbs measure on $\O_n=\{-1, 1\}^{\L_n}$ with boundary condition
$\s_n^{(+)}$ is defined in the usual way

$$\m^+_{n,\b}(\s)=Z^{-1}(n,\b,+)
\exp(-\b H_n^+(\s)), \eqno(3)$$
where $\b=T^{-1}$, $T>0-$ temperature and
 $ Z(n, \b, +)$ is the normalizing factor (statistical sum).

Denote by $\s^+_n$ the configuration on $\Z$ such that
$\s^+_n(x)\equiv +1$ for any $x\in \L_n^c.$

Put
$$A(\s^+_n)=\{x\in \Z: \s^+_n(x)=-1\},$$
$$\d(\s^+_n)=\{l=(m-1,m)\in \Z\times \Z: \s^+_n(m-1)\ne \s^+_n(m)\}.$$
Note that there is one-to-one correspondence between the set
of all configurations
$\s^+_n$ and the set of all subsets $A$ of $\L_n.$

Let $A'(\s^+_n)$ be the set of all maximal connected subsets of
$A(\s^+_n).$

\vspace{0.3cm}
 LEMMA 1. {\it Let $B\subset \Z$ be a fixed connected set and
$p_\b^+(B)=\mu^+_{n,\b}\{\s^+_n: B\in A'(\s^+_n)\}.$ Then
$$p^+_\b(B)\leq \exp\bigg\{-\b\bigg[I_{n_B}+I_{N_B+1}\bigg]\bigg\}, $$
where $n_B$ (resp. $N_B$) is the left (resp. right)  endpoint of $B.$}
\vspace{0.3cm}

{\it Proof.} Denote $F_{B}=\{\s^+_n: B\in A'(\s^+_n)\}-$ the set of all configurations $\s^+_n$ on $Z$ with
"+"-boundary condition (i.e. $\s^+_n(x)\equiv +1$ for any $x\in \L_n^c$) such that $B$ is maximal connected set.
Denote also $F_{B}^-=\{\s^+_n: B\cap A'(\s^+_n)=\emptyset\}.$ Define the map $\chi_{B}:F_{B}\to F_{B}^-$ as
following: for $\s_n\in F_{B}$ we destroy the set $B$ changing the values $\s_n(x)$ inside of $B$ to $+1$. The
constructed configuration is $\chi_{B}(\s_n)\in F^-_{B}.$

For a given $B$  the map
$\chi_{B}$ is one-to-one map.

For $\s_n\in F_B$ it is clear that
$$A'(\s_n)=A'(\chi_{B}(\s_n))\cup B , \ \
\d (\s_n)=\d(\chi_{B}(\s_n))\cup \{(n_B-1, n_B), (N_B, N_B+1)\}.$$

Thus we have
$$H^+_n(\s_n)-H^+_n(\chi_B(\s_n))=I_{n_B}+I_{N_B+1}.\eqno(4)$$

By definition we have
$$p_\b^+(B)={\sum\limits_{\s_n \in  F_B}\exp\{-\b H_n^+(\s_n)\}\over
\sum\limits_{\s_n}\exp\{-\b H_n^+(\s_n)\}}\leq   {\sum\limits_{\s_n\in F_B}\exp\{-\b H_n^+(\s_n)\}\over
\sum\limits_{\s_n\in F_B}\exp\{-\b H_n^+(\chi_B(\s_n))\}}. \eqno(5)$$ Using (4) from (5) we get

$$p_\b^+(B)\leq {\sum\limits_{\s_n \in  F_B}\exp
\bigg\{-\b H^+_n(\chi_B(\s_n))-\b\bigg[I_{n_B}+I_{N_B+1}\bigg]\bigg\}
\over \sum\limits_{\s_n\in F_B}\exp\{-\b H_n^+(\chi_B(\s_n))\}}=$$
$$\exp\bigg\{-\b\bigg[I_{n_B}+I_{N_B+1}\bigg]\bigg\}.$$
The lemma is proved.

Assume that for any $r\in \{1,2,...\}$ and $n\in \Z$ the Hamiltonian (1)
satisfies the following condition
$$ I_n+I_{n+r}\geq r. \eqno(6)$$

\vspace{0.3cm}
LEMMA  2. {\it Assume condition (6) is satisfied.
Then  for all sufficiently large $\b$, there is a
constant $C=C(\b)>0,$  such that
$$\m^+_\b\{\s_n: \ \ |B|>C\ln |\L_n|\ \ \mbox{for some} \ \
B\in A'(\s_n)\}\to 0,\ \ \mbox{as} \ \  |\L_n |\to \infty ,$$ where
$|\cdot|$ denotes the number of elements.} \vspace{0.3cm}

{\it Proof.}
Suppose $\b> 1$, then by Lemma 1 and condition (6) we have
$$\m^+_{\b}\{\s_n: B\in A'(\s_n), \ \ t\in B, |B|=r\}=
\sum_{B: \ \ t\in B, \ \ |B|=r}p^+_\b(B)\leq r\exp\{-\b r\}.$$

Hence
$$\m^+_{\b}\{\s_n: B\in A'(\s_n), \ \ t\in B, |B|>C_1\ln |\L_n|\}
\leq \sum_{r\geq C_1\ln |\L_n|} r\exp\{-\b r\}\leq $$
$$\sum_{r\geq C_1\ln |\L_n|} \exp\{(1-\b) r\}={|\L_n|^{C_1(1-\b)}
\over 1-e^{1-\b}}, \eqno(7)$$
where $C_1$ will be defined latter. Thus we have
$$\m^+_{\b}\{\s_n: \exists B\in A'(\s_n), \ \  |B|>C_1\ln |\L_n|\}
\leq {|\L_n|^{C_1(1-\b)+1}
\over 1-e^{1-\b}}.$$

The last expression tends to zero if $|\L_n|\to\infty$ and $C_1>{1\over \b-1}.$
The lemma is proved.
\vspace{0.3cm}

LEMMA 3. {\it  Assume the condition (6) is satisfied.
Then
$$\m^+_{\b}\{\s_n: \s_n(0)=-1\}\to 0, \ \ \mbox{as} \ \
 \b\to \infty.\eqno (8)$$ }

\vspace{0.3cm}
{\it Proof.} If $\s_n(0)=-1,$ then $0$ is point for
 some $B\in A'(\s_n).$
Consequently,
$$\m^+_{\b}\{\s_n: 0\in B, \ \  |B|<C_1\ln|\L_n|\}
\leq \sum^{C_1\ln|\L_n|}_{r=1}(e^{1-\b})^r\leq
{e^{1-\b}\over 1-e^{1-\b}}$$ and
$$\m^+_{\b}\{\s_n(0)=-1\}\leq \m^+_{\b}\{\s_n: 0\in B,
\ \  B \in A'(\s_n)\}\leq
$$
$$
{e^{1-\b}\over 1-e^{1-\b}}+
{|\L_n|^{C_1(1-\b)+1}
\over 1-e^{1-\b}}.\eqno (9)$$

For $|\L_n|\to \infty$ and $\b\to\infty$ from (9) we get
(8). The lemma is proved.

\vspace{0.3cm}
 THEOREM 4. {\it Assume the condition (6) is satisfied.
 For all sufficiently large $\b$ there are at least
two Gibbs measures for the  model (1) .}
\vspace{0.3cm}

{\it Proof.}  Using a similar argument one can prove
$$\m^-_{\b}\{\s_n: \s_n(0)=1\}\to 0, \ \ \mbox{as} \ \
 \b\to \infty.$$
Consequently, for sufficiently large $\b$ we have
$$\m^+_{\b}\{\s_n: \s_n(0)=-1\}\ne \m^-_{\b}\{\s_n: \s_n(0)=-1\}.$$
This completes the proof.

Denote
$${\cal H}=\{H: H \mbox{ (see (1)) satisfies the condition (6)}\}$$
The following example shows that the set ${\cal H}$ is not empty.

\vspace{0.3cm}
{\bf Example.} Consider Hamiltonian  (1) with
$I_m\geq |m|,\ \  m\in \Z.$
Then
$$I_m+I_{m+k}\geq |m|+|m+k|\geq k$$
for all $m\in \Z$ and $k\geq 1.$
Thus the condition (6) is satisfied.

\vspace{0.3cm} {\bf Acknowledgments.} The work supported by NATO Reintegration Grant : FEL. RIG. 980771. The final
part of this work was done in the Physics Department of ``La Sapienza'' University in Rome. I thank M.Cassandro
and G.Gallavotti for an invitation to the ``La Sapienza'' University and useful discussions. I thank the referee
for useful sugestions.

\vskip 0.3 truecm
{\bf References}

1. Baur M., Nosanow L.
{\it J. Chemical Phys.} {\bf 37}: 153-160 (1962).

2. Cassandro M., Ferrari P., Merola I., Presutti E.
{\it arXiv: math-ph/ }021 1062, {\bf 2} 28 Nov 2002.

3. Dyson F.  {\it Commun. Math. Phys.} {\bf 12}: 91-107 (1969)

4. Dyson F.
{\it Commun. Math. Phys.} {\bf 21}: 269-283 (1971)

5. Flohlich J., Spencer T.
{\it Commun. Math. Phys.} {\bf 84}: 87-101 (1982)

6. Johansson K.
{\it Commun. Math. Phys.} {\bf 141}: 41-61 (1991)

7. Johansson K.
{\it Commun. Math. Phys.} {\bf 169} : 521-561 (1995)

8. Little W.,
{\it Phys. Rev.} {\bf 134}: A1416-A1424 (1964)

9. Ruelle D.,  {\it Statistical mechanics (rigorous results)},
Benjamin, New York, 1969

10. Strecker J. {\it J. Math. Phys.}
{\bf 10}: 1541-1554 (1969)

11. Van Hove L. {\it Physica} {\bf 16}: 137-143 (1950).

\end{document}